\documentclass[11pt,twoside]{article}

\usepackage{asp2006}
\usepackage{epsf}
\usepackage{lscape}

\begin{document}
\title{Protoplanetary disks of  TTauri binaries in Orion: Prospects for planet formation}   %%% Fill in title
\author{M. G. Petr-Gotzens and S. Daemgen}   %%% Fill in author names
\affil{European Southern Observatory, Karl-Schwarzschild-Str.\ 2, D-85748 Garching, Germnay}    %%% Fill in author affiliations
\author{S. Correia}
\affil{Astronomisches Institut Potsdam, Schillerg\"asschen, Potsdam, Germany}

\begin{abstract} %%% Abstract to run on from here.
Dusty protoplanetary disks surrounding young low-mass stars are the 
birthplaces of planets. Studies of the evolutionary timescales of such disks
provide important constraints on the timescales of planet formation. Binary 
companions, however, can influence circumstellar disk evolution through 
tidal interactions.
In order to trace protoplanetary disks and their properties in young binary systems,
as well as to study the effect of binarity on circumstellar disk lifetimes,
we have carried out spatially resolved spectroscopy for several low-mass
binaries in the well-known Orion Nebula Cluster. Br$_{\gamma}$ emission, which we 
detect in several systems, is used as a tracer
for the presence of an active accretion disk around a binary component. 
We find a paucity of actively accreting secondaries, and hence, evidence that in a binary system it is 
the lower mass component that disperses its disk faster. 
\end{abstract}

%%% MAIN BODY OF TEXT GOES HERE. CONSULT "INSTRUCTIONS FOR AUTHORS USING
%%% LATEX2E MARKUP", SECTIONS 2.3-2.6 FOR HELP WITH EQUATIONS, FIGURES,
%%% AND TABLES.

\section{Introduction}   %%% Top level section head (remove "%" symbol)
Circumstellar proto-planetary disks are not only an inevitable product of the low-mass star 
formation process,  but they are also the most important pre-requisite for the formation of life.  It is the 
dust and gas content of such disks out of which planetary systems are born. Therefore, investigations
of  the composition and evolution of 
proto-planetary disks around young low-mass stars, is directly linked to our understanding of planet 
formation, and ultimately elucidating the process that has been responsible for the formation of our own
solar system. 

The overall lifetime of circumstellar disks, and hence the upper limit on planet-building timescales, 
has been determined to approximately 10-6Myr for stars in low stellar density regions and stellar clusters
respectively \citep{cieza2007,carp2006,haisch2001}. These findings are based on infrared observations 
that measure, for stellar groups of different age, the fraction of stars with circumstellar disks as identified by
their near- and mid-infrared excess. One of the dominant processes driving disk evolution in single stars
seems to be photoevaporation from the central stellar source \citep{cieza2008}, suggesting that
more luminous stars loose their disks faster.  Indeed, studies of large samples of roughly coeval stars
spanning a range in mass from $\sim 0.1{\rm M}_{\odot}$ to $\sim 7{\rm M}_{\odot}$ found that lower mass stars
retain their disks for a longer time than higher mass stars \citep{kenken2009}. 

On the other hand, many stars are members of a binary or  multiple system, and for nearby solar-like stars the
binary fraction is even as high as $\sim 60\%$ \citep[e.g][]{duq1991}.
Hence tidal truncation is expected to be an additional important 
parameter that should govern the lifetime of a circumstellar disk around each individual binary component.
Theoretically, one expects that the truncation radius of the outer circumstellar disk scales with the binary separation
and with the components' mass as $R_t  \sim 0.3-0.5 \times a$, with $a$ being the binary separation and its
factor varying with the system's mass ratio \citep{armin1999, papa1977}. 
Because the truncation of the disks should limit the amount of disk
material that can be accreted, reduced disk lifetimes are expected for binaries, and in particular for
systems with smaller separations, and for {\it less massive} stars.  The presence of a stellar component
may also lead to shorter disk accretion timescales for stars in binary systems
as compared to single stars.

\section{Goal of this study}

The goal of this study is to determine the presence of a circumstellar protoplanetary
disk around each individual component
of a sample of close binary stars in the $\sim$1Myr old Orion Nebula Cluster. Thereby, we wish
to investigate if, and how, the evolutionary timescales of disks around young stars being members of
a binary depart from those determined for single stars. The following diagnostics 
have proven to be very good indicators for the presence of a disk: (i) photometric 
excess emission at K- and L-band reliably 
traces the presence of a warm inner disk \citep[e.g.][]{cabe2006}, and
(ii)  Br$_{\gamma}$ emission at 2.16$\mu$µm indicates magnetospheric accretion from a circumstellar
disk onto the central young star \citep[e.g.][]{muzz2001} . 

In this contribution
we focus on spectroscopic observations only and discuss our search for Br$_{\gamma}$ emission 
in spatially resolved, low-mass ($<2 {\rm M}_{\odot}$) binaries located in the Orion Nebula 
Cluster (ONC).
At a distance of $\sim$440pc the ONC is the closest region of active low- and high-mass star 
formation. Most of the stars in the field of our Galaxy, and most likely our sun as well, were formed in 
OB star clusters as those present in Orion. 

\section{Observations and data reduction}
A total of 22 ONC binaries have been observed with K-band adaptive 
optics spectroscopy at Gemini Observatory, using NIFS (spectral resolution of R$\sim$5000),
or at ESO's VLT, using NACO (R$\sim$1400). Both instruments provide a wavelength coverage of $\sim 2.05-2.45 \mu$m.
The larger data subset (16 targets) was obtained with the NACO instrument. The range of separation of
the binary components is $0.26^{\prime\prime}-1.1^{\prime\prime}$, which corresponds to about 110-500\,AU
at the distance of the ONC.

 \begin{figure}[!ht] 
%\plotfiddle{JW876_posterplot_b.eps}{vsize=8cm]{rot=90}{}{}{}{}
\epsscale{0.8}
\plotone{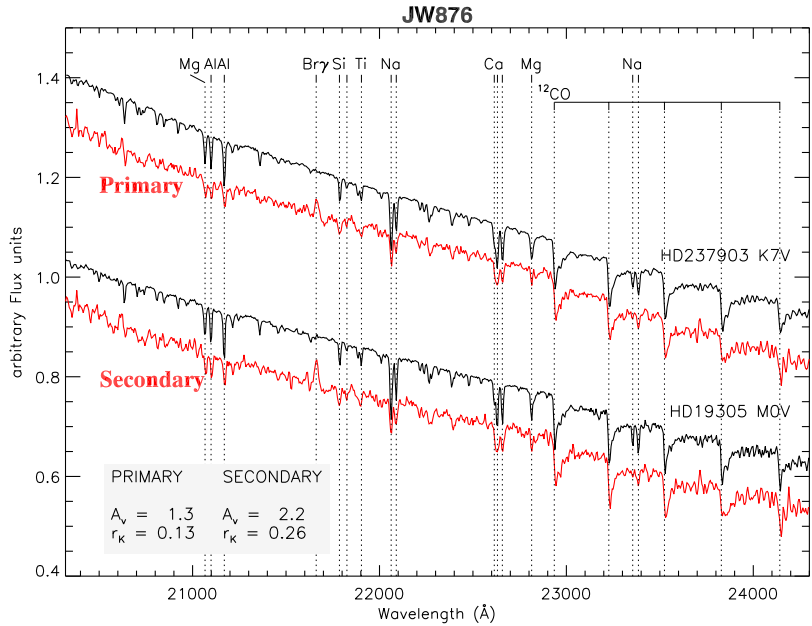}
\caption{Spectrum of the primary and secondary component of the $0.5^{\prime\prime}$ binary JW876. Each 
spectrum is plotted together with a template dwarf spectrum from the IRTF spectral library. Both components
clearly show Br$_{\gamma}$ in emission.}
\label{fig1}
\end{figure}

After standard data reduction of the 2-dim spectral images (flat fielding, sky subtraction, bad pixel correction) 
IRAF/apextract was employed to extract the spectrum of the primary and secondary of each 
binary. The spectra were further cleaned from telluric lines through division by standard star spectra of spectral
types B0-B9V, 
which had been observed close in airmass and time. Before the division, Br$_{\gamma}$ absorption in the standard star
spectra was removed by interpolation, allowing us to probe all science target spectra for Br$_{\gamma}$-emission,
that is indicative for active accretion. Furthermore,
spectral typing was achieved by matching the spectra with templates from the IRTF Spectral 
Library \citep[Rayner et al. 2009 in prep.,][]{cush2005}. To find the best template, as well as the 
corresponding best reddening and veiling/excess values, a range of A$_V$ and K-band excess were 
applied to all spectral templates. Then, a minimum chi$^2$ was used to fit the modified templates with 
the science spectra. As an example, we show in Figure~\ref{fig1} the result for the binary JW876.

\section{First Results}
\begin{figure}[!ht] 
\epsscale{0.8}
\plotone{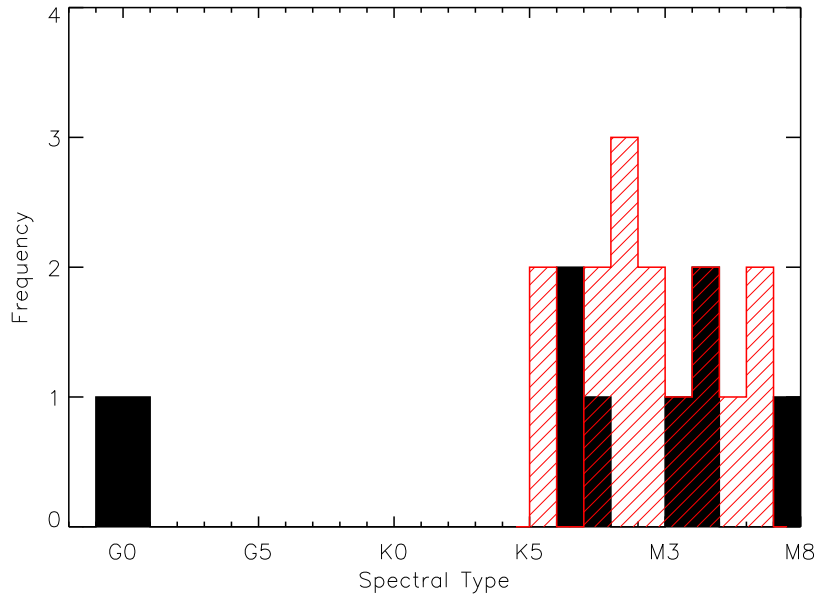}
\caption{Distribution of spectral types for stellar components that do show Br$_{\gamma}$ emission (solid
bars) and that do not show Br$_{\gamma}$ emission in their spectrum (hatched bars).}
\label{fig2}
\end{figure}

The first evaluation of the reduced spectra provided spectral types for almost all binary components. The large 
majority of the stars are of K and M spectral type, as expected for young 
low-mass objects at the age of 1 Myr. Our preliminary analysis suggests
Br$_{\gamma}$-emission lines in several of the binary components. In detail, we find that a fraction of 
$\sim80 \%$ of the target systems show any sign of an active accretion disk, i.e. most of the binaries
possess at least one component with a clear signature of Br$_{\gamma}$-emission. The numbers  
of pairs where both components show Br$_{\gamma}$-emission and where only one component shows
Br$_{\gamma}$ (so-called mixed-pairs) are roughly equal, and we conclude that mixed-pairs are 
common. Quite intriguingly, only one system was found in which {\it only} the secondary seems to have 
an active accretion disk. This finding is apparently not due to an observational bias 
in the sense that 
the detection of the Br$_{\gamma}$ emission line is more difficult in secondaries, because they 
are usually the less massive and later spectral type component: in Figure~\ref{fig2} we show that
for those systems observed in this study the 
presence of Br$_{\gamma}$ emission is not a function of spectral type. 
It is evident from the data, although not yet statistically significant, that 
binarity influences disk evolution, even for binaries as wide as our ONC targets, all of which have 
separations $>$100 AU. The under-representation of active accretion disks among secondaries 
hints at disk dissipation working faster on (potentially) lower mass secondaries, leading us to 
speculate that secondaries are possibly less likely to form planets. A similar result has been 
reported by \citet{whighez2001}.

In the context of the latter, we further note an interesting observational result: Almost 40 of all the extra-solar planets 
discovered to date reside in wide binary systems where the component separation is larger than 
100AU (large 
enough that planet formation around one star should not strongly be inßuenced by the companion star). But for all 
these systems an extra-solar giant planet is found only around one star, never around both stars of the binary 
system \citep[see][]{desi2007}, and in 95\% of the cases it is the primary component that hosts
the extra-solar planet. 

%planet formation would still be possible in the circumstellar disks of binary secondaries (e.g. Lissauer et al. 2004, 
%Quintana et al. 2005, 2007), in particular in systems having separations of >100AU, the canonical disk size. 
%%Circumstellar disk evolution of components in such wide binary system is thought to be not affected by the 
%companion (e.g. Armitage et al. 2003, Pascucci et al. 2008). 

%\subsection{}   %%% Second level section head (remove "%" symbol)
%\subsubsection{}   %%% Lowest level section head (remove "%" symbol)
%\section*{}    %%% Unnumbered top level section head (remove "%" symbol)
%\subsection*{}   %%% Unnumbered second level section head (remove "%" symbol)

\acknowledgements %%% Text of acknowledgements runs on after this command.
We thank the ESO La Silla/Paranal Observatory staff for their great support in carrying out 
our scientific programme in service observing mode. 

%%% THE BIBLIOGRAPHY
%%%
%%% CONSULT SECTION 3 OF "INSTRUCTIONS FOR AUTHORS" FOR HOW TO USE NATBIB.
%%% AUTHORS ARE ENCOURAGED TO USE EITHER THE "THEBIBLIOGRAPY" ENVIRONMENT
%%% BY UNCOMMENTING (DELETING THE "%" SYMBOL) THE COMMANDS BELOW, OR BY
%%% USING THE BIBTEX ENVIRONMENT. TO FIND OUT WHICH IS APPLICABLE TO YOUR
%%% CONTRIBUTION, CONSULT THE VOLUME EDITORS FOR YOUR PROCEEDINGS.
%%%

\end{document}